\def\rfr#1{eq. (\ref{#1})}

\def\dert#1#2{\frac{{{d}}{#1}}{{{d}}{#2}}}              

\def\virg#1{``#1''}

\def\eqi{\begin{equation}}
\def\eqf{\end{equation}}
\def\eqia{\begin{eqnarray}}
\def\eqfa{\end{eqnarray}}
\def\dmdt{\left(\rp{\dot m}{m}\right)}
\def\vesc{V_{\rm esc}}
\def\rp#1#2{{#1\over#2}} \def\lb#1{\label{#1}}

\def\bds#1{\boldsymbol{#1}}

\documentclass{elsart}

\usepackage{slashbox}
\usepackage{url}
\usepackage{amsmath}
\usepackage{latexsym,wasysym}

\usepackage{natbib}

\usepackage{epsfig}

\usepackage{amssymb}

\RequirePackage{color}

\begin{document}

\begin{frontmatter}



\title{Orbital effects of  non-isotropic mass depletion of the atmospheres of evaporating hot Jupiters \textcolor{black}{in extrasolar systems}}


\author{Lorenzo Iorio\thanksref{footnote2}}
\address{Ministero dell'Istruzione, dell'Universit\`{a} e della Ricerca (M.I.U.R.)-Istruzione\\
International Institute for Theoretical Physics and
High Mathematics Einstein-Galilei\\ Fellow of the Royal Astronomical Society (F.R.A.S.)}
\thanks[footnote2]{Address for correspondence:  Viale Unit\`{a} di Italia 68, 70125, Bari (BA), Italy}
\ead{lorenzo.iorio@libero.it}
\ead[url]{http://digilander.libero.it/lorri/homepage$\_$of$\_$lorenzo$\_$iorio.htm}



\begin{abstract}
We analytically and numerically investigate the  {long-term}, i.e. averaged over one full revolution, orbital effects of the non-isotropic percent mass loss $\dot m/m$ experienced by several transiting hot Jupiters whose atmospheres are hit by severe radiations flows coming from their close parent stars. The semi-major axis $a$, the argument of pericenter $\omega$ and the mean anomaly $\mathcal{M}$ experience net  variations, while the eccentricity $e$, the inclination $I$ and the longitude of the ascending node $\Omega$ remain unchanged {, on average}. In particular, $a$ increases independently of $e$ and of the speed $V_{\rm esc}$ of the ejected mass. By assuming $ {|\dot m|}\lesssim 10^{17}$ kg yr$^{-1}$,  {corresponding to $|\dot m/m|\lesssim 10^{-10}$ yr$^{-1}$ for a Jupiter-like planet}, it turns out $\dot a\sim 2.5$ m yr$^{-1}$ for orbits with $a=0.05$ au. Such an effect may play a role in the dynamical history of the hot Jupiters, especially in connection with the still unresolved  issue of the arrest of the planetary inward migrations after a distance  $a\gtrsim 0.01$ au is reached. The retrograde pericenter  {variation} depends, instead, on $e$ and $V_{\rm esc}$. It may, in principle, act as a source of systematic uncertainty in some proposed measurements of the general relativistic pericenter precession; however, it turns out to be smaller than it by several orders of magnitude.
\end{abstract}


\end{frontmatter}

\parindent=0.5 cm

\section{Introduction}
As shown first by\footnote{See also \citet{Somm}, \citet{russa}, \citet{Hadji1} and the review by  \citet{Hadji}. } \citet{Mesc}, the equation of motion for a body acquiring or ejecting mass due to interactions with its environment  is\footnote{It is just the case to remind that an erroneous form of \rfr{equazza}, in which $-\bds v$ appears instead of $\bds u$, was not rarely used in literature. See, e.g., \citet{erro} for a discussion of such a topic. Suffice it to say that it would be non-invariant under Galilean transformations.}
\eqi \dert{\bds v} t = \rp{\bds F}{m} + \dmdt\bds u.\lb{equazza}\eqf
with respect to some inertial frame $K$.
In it, $\dot m\doteq dm/dt$, $\bds F$ is the sum of all the external forces, and
\eqi\bds u\doteq \bds{v}_{\rm esc}-\bds v\lb{velrela}\eqf
is the velocity of the escaping mass with respect to the barycenter  of the body: $\bds v$ is the velocity of\footnote{ {More precisely, it is the velocity of that point of the body which, at each moment, coincides with the body's center of mass. It does not include the geometric shift of the center of mass caused by the mass loss.}} the center of mass of the body with respect to $K$, while\footnote{\textcolor{black}{To avoid confusions, it may be useful to remark that in literature $\bds{v}_{\rm esc}$ is sometimes denoted by $\bds u$, while no specific notations are used for the relative velocity in \rfr{velrela}. Cfr., e.g., \citet{russa}}.} $\bds{v}_{\rm esc}$ is the velocity of the escaping particle with respect to $K$. If the mass loss is isotropic with respect to the body's barycenter, then the total contribution\footnote{ {Strictly speaking, it is to be intended as $(\sum_i \dot m_i\bds u_i)/m$, where the label $i$ refers to the $i$th point of the surface from which mass is escaping.}} $\left(\dot m/m\right)\bds u$ vanishes. Notice that \rfr{equazza} fulfils the Galilean invariance under a transformation $\bds v\rightarrow\bds v^{'}=\bds v-\bds V$ to another inertial frame $K^{'}$ moving with constant velocity $\bds V$ with respect to $K$.

An interesting,  {real} physical scenario in which \rfr{equazza} is applicable is represented by those transiting exoplanets whose atmospheres are escaping\footnote{See http://www.spacetelescope.org/news/heic0403/ on the WEB.}  because of the severe levels of energetic radiations, coming from their very close parent stars, hitting them \citep{trans}. For those few transiting planets
which are observed in the ultraviolet it is possible to gain information about the size and mass-loss rate
of their evaporating upper atmospheres. It turns out that typical values of the evaporation rate for HD 209458b \citep{Char00,Hen00}, unofficially named \virg{Osiris}, HD 189733b \citep{Bou05}, and WASP-12b \citep{Hebb09} are in the range \citep{eren,Foss,Lins,Leca,vida03,vida04,vida08} \eqi\left|\dot m\right|\sim 3\times 10^{14}-10^{15}\ {\rm kg\ yr}^{-1}=10^{10}-10^{11}\ {\rm g\ s}^{-1}\lb{dmd};\eqf however, \citet{trans} estimate the mass-loss rates of all detected transiting planets finding an upper bound as large as
\eqi\left|\dot m\right|\lesssim 3\times 10^{17}\ {\rm kg\ yr}^{-1}=10^{13}\ {\rm g\ s}^{-1}.\lb{fluffo}\eqf Several, still unresolved issues are connected with giant exoplanets like the aforementioned ones. Indeed, they all orbit at less than $0.1$ au from their parent stars: thus, they certainly could not have formed there, so that an inward migration from more distant locations should have occurred \citep{Lin96,Lubo}. Why did such hot Jupiters stop migrating at such distances? What was the fate of  those gaseous giants that migrated further in, if any? Is the evaporation responsible of the apparent lacking of exoplanets closer than $0.01$ au to their host stars?

Studying the orbital consequences of \rfr{equazza} may help to  shed some light on such important open problems.
In Section \ref{due} we work out, both analytically and numerically, the long-term variations of all the standard six Keplerian orbital elements induced by \rfr{equazza}. Then, we apply the results obtained to some specific exoplanetary scenarios. Section \ref{tre} summarizes our findings.
\section{Orbital effects}\lb{due}
In the case of an evaporating transiting exoplanet  orbiting a close Sun-like star, $\bds F$ is the usual Newtonian gravitational monopole.
Let us write down \rfr{equazza} for both the planet p, with mass $m$, and its hosting star s, with mass $M$, with respect to some inertial frame $K$ \textcolor{black}{\citep{russa}}:
\begin{equation}
\left\{
\begin{array}{lll}
\dert{{\bds v}_{\rm p}} t & = & \rp{{\bds F}_{\rm p}}{m} + \dmdt\bds u_{\rm p}, \\ \\
\dert{{\bds v}_{\rm s}} t & = & \rp{{\bds F}_{\rm s}}{M} + \left(\rp{\dot M}{M}\right)\bds u_{\rm s},
\end{array}
\right.\lb{orbi}
\end{equation}
where
\begin{equation}
\left\{
\begin{array}{lll}
\bds F_{\rm p}& = &-\rp{GMm}{r^3}\left(\bds r_{\rm p}-\bds r_{\rm s}\right), \\ \\
\bds F_{\rm s}& = &-\rp{GMm}{r^3}\left(\bds r_{\rm s}-\bds r_{\rm p}\right),
\end{array}\lb{New}
\right.
\end{equation}
and
\eqi r\doteq\left|\bds r_{\rm p}-\bds r_{\rm s}\right|.\eqf
 In \rfr{orbi} we took into account the possibility that also the star experiences a mass loss due to internal physical processes. In the case of the Sun,  it is estimated to be of the order of \citep{MASLOS}
\eqi\rp{\dot M}{M}=-9\times 10^{-14}\ {\rm yr}^{-1};\lb{pazzia}\eqf
about 80$\%$ of such a mass-loss is due to the core nuclear burning, while the remaining $20\%$ is due to average solar wind. In any case, it can be considered isotropic, so that it is $(\dot M/M)\bds u_{\rm s}=0$. Actually, most of the existing literature  {(see, e.g., \citet{gyl,strom,jeans,armel,vescan,vescan2,jeans61,Hadji1,omarov,had,kevo,khol,verh,kury,deprit,polya,prietoa,prietob,Li03,Li08,Li09,rahoma,Iorio})} is devoted on treating the motion in a two-body system with various kinds of time-dependent isotropic mass loss  {affecting the Newtonian monopole itself; for a recent treatment of such a topic with the same method of the present paper and a discussion of some of the approaches present in literature, see \citet{Iorio}}.
Concerning the evaporating planet,  let us assume for it a constant percent decrement $\dot m/m$ of its mass; it is  especially true over timescales of the order of one orbital period $P_{\rm b}$, if the orbit is assumed almost circular\footnote{It should be not the case \citep{iro,trans} for, e.g., HD 80606b \citep{naef} whose orbit has an eccentricity as large as $e=0.93$. Recall that it is a numerical parameter determining the shape of the Keplerian ellipse: $0\leq e<1$, with $e=0$ corresponding to a circle.}. For a typical Jovian mass $m\sim m_{\rm J}=1.899\times 10^{27}$ kg and the figure of \rfr{fluffo} we have
\eqi\left|\dot m/m\right|\lesssim 1.7\times 10^{-10}\ {\rm yr}^{-1}.\lb{dmd2}\eqf Moreover, its mass decrement is clearly non-isotropic with respect to its center of mass, so that $(\dot m/m)\bds u_{\rm p}$ does not vanish. As a result,  from \rfr{orbi} and \rfr{New} it can be obtained the equation for the relative planet-star motion\footnote{\textcolor{black}{Cfr. with the classification in Table I of \citet{russa}: \rfr{equazza2} falls within the A.II.b.9 (Seeliger) or B.II.b.20 (Fermi) cases depending on the relative sizes of $m$ and $M$.}} \textcolor{black}{\citep{russa}}
\eqi\dert{\bds v} t  =  -\rp{\mu}{r^3}\bds r + \dmdt\bds u_{\rm p},\lb{equazza2}\eqf
where we defined
\eqi\mu\doteq G(M+m),\ \bds r\doteq \bds r_{\rm p}-\bds r_{\rm s},\ \bds v\doteq \bds v_{\rm p}-\bds v_{\rm s}.\eqf
Moreover, $\bds u_{\rm p}$ can be conveniently expressed as the difference between the escape velocity $\bds V_{\rm esc}$ with respect to the star and the velocity $\bds v$ of the planet with respect to the star.
Clearly, $\bds V_{\rm esc}$ is radially directed from the parent star to the planet, i.e.
\eqi\bds V_{\rm esc}=V_{\rm esc}\bds{\hat{R}},\ \bds{\hat{R}}\doteq \rp{\bds r_{\rm p}-\bds r_{\rm s}}{r},\eqf and\footnote{Actually, such a figure comes from the escape speed from the planet $q_{\rm esc}=\sqrt{2Gm/R_{\rm p}}$: in order to have $V_{\rm esc}$ one should also add  the radial component of the planet's motion which, however, is negligible because of the assumed low eccentricity of its orbit (see below, \rfr{velo}). } \citep{vida04} \eqi V_{\rm esc}\sim 10^4\ {\rm m\ s}^{-1}.\eqf
In order to evaluate  the magnitude of $\bds u_{\rm p}$, let us note that in most of the considered exoplanets the orbits are almost circular, so that it can be posed
\eqi u_{\rm p}=\sqrt{v^2+V_{\rm esc}^2-2\left(\bds v\bds\cdot\bds V_{\rm esc}\right)}\sim\sqrt{v^2+V_{\rm esc}^2};\eqf by assuming that a Jupiter-sized planet is at a distance\footnote{Here $a$ is the semi-major axis of the Keplerian ellipse: it defines its size.} $a$ of the order of $0.05$ au from its Sun-like parent star, it turns out
\eqi v\sim \sqrt{\rp{\mu}{a}}\sim 10^5\ {\rm m\ s}^{-1},\eqf so that
\eqi u_{\rm p}\sim v\sim 10^5\ {\rm m\ s}^{-1}.\lb{uvel}\eqf

Such figures imply that the mass-escaping term in \rfr{equazza2} can be considered as a small perturbation $A_{\dot m}$ with respect to the Newtonian monopole $A_{\rm N}$;
indeed, from \rfr{dmd2} and \rfr{uvel} turns out
\begin{equation}
\left\{
\begin{array}{lll}
A_{\rm N} & \sim & 2\ {\rm m\ s}^{-2}, \\ \\
A_{\dot m}& \sim & 5\times 10^{-13}\ {\rm m\ s}^{-2}.
\end{array}
\right.
\end{equation}
Thus, it can be treated with the standard perturbative techniques like, e.g., the Gauss equations  for the variation of the  {osculating} Keplerian orbital elements \citep{Brou} which are valid for any kind of disturbing acceleration $\bds A$, irrespectively  of its physical origin. \textcolor{black}{See also \citet{omarov62} for their use in the variable mass case.}
 {The Gauss equations are
\begin{equation}
\begin{array}{lll}
\dert a t & = & \rp{2}{n\sqrt{1-e^2}} \left[e A_R\sin f +A_{T}\left(\rp{p}{r}\right)\right],\\   \\
\dert e t  & = & \rp{\sqrt{1-e^2}}{na}\left\{A_R\sin f + A_{T}\left[\cos f + \rp{1}{e}\left(1 - \rp{r}{a}\right)\right]\right\},\\  \\
\dert I t & = & \rp{1}{na\sqrt{1-e^2}}A_N\left(\rp{r}{a}\right)\cos u,\\   \\
\dert\Omega t & = & \rp{1}{na\sin I\sqrt{1-e^2}}A_N\left(\rp{r}{a}\right)\sin u,\\    \\
\dert\varpi t & = &\rp{\sqrt{1-e^2}}{nae}\left[-A_R\cos f + A_{T}\left(1+\rp{r}{p}\right)\sin f\right]+2\sin^2\left(\rp{I}{2}\right)\dert\Omega t,\\   \\
\dert {\mathcal{M}} t & = & n - \rp{2}{na} A_R\left(\rp{r}{a}\right) -\sqrt{1-e^2}\left(\dert\omega t + \cos I \dert\Omega t\right).
\end{array}\lb{Gausseq}
\end{equation}
}
  {In \rfr{Gausseq} $n\doteq\sqrt{\mu/a^3}$ is the unperturbed Keplerian mean motion, $e$ is the eccentricity,  $p\doteq a(1-e^2)$ is the semi-latus rectum, $I$ is\footnote{The parameters $I,\Omega,\omega$ defines the spatial orientation of the Keplerian ellipse, which, in the unperturbed case, changes neither its size nor its shape: they can be thought as the three Euler angles fixing the orientation of a rigid body in the inertial space.} the inclination of the orbital plane to the reference $\{x,y\}$ plane chosen, $\Omega$ is the longitude of the ascending node, $\omega$ is the argument of the pericenter, $\mathcal{M}$ is the mean anomaly\footnote{It is connected with the time of pericenter passage $t_p$ through $\mathcal{M}\doteq n(t-t_p)$.}, $f$ is the true anomaly, $u\doteq \omega+f$ is the argument of latitude, and $A_R,A_T,A_N$ are the radial, transverse and out-of-plane components of the disturbing acceleration $\bds A$, respectively.}
To this aim, $\bds A_{\dot m}$ can be written
\eqi \bds A_{\dot m} = \dmdt\left[(V_{\rm esc}-v_R)\bds{\hat{R}}-v_{T}\bds{\hat{T}}\right],\lb{apert}\eqf
where $v_R,v_{T}$ are the\footnote{Here  $\bds{\hat{T}}$ denotes the unit vector along the  transverse direction, i.e. orthogonal to $\bds{\hat{R}}$.} radial and transverse components of the planet's velocity  evaluated onto the unperturbed Keplerian ellipse
\begin{equation}
\left\{
\begin{array}{lll}
v_R & = & \rp{nae\sin E}{1-e\cos E}, \\ \\
v_{T} & = & \rp{na\sqrt{1-e^2}}{1-e\cos E},
\end{array}
\right.\lb{velo}
\end{equation}
In \rfr{velo},  $E$ is the eccentric anomaly\footnote{It can be regarded as
a parametrization of the usual polar angle $\theta$ in the orbital plane.}.
 From \rfr{apert} it is inferred that $A_N=0$, being  $A_R$ and $A_T$ the only non-zero components of $\bds A_{\dot m}$;
\textcolor{black}{
they are
\begin{equation}
\left\{
\begin{array}{lll}
A_R & = & \left(\rp{\dot m}{m}\right)\left(V_{\rm esc}-\rp{nae\sin E}{1-e\cos E}\right), \\ \\
A_T & = & -\left(\rp{\dot m}{m}\right)\rp{na\sqrt{1-e^2}}{1-e\cos E}.
\end{array}
\right.\lb{accli}
\end{equation}
}

The  {long-term} effects of \rfr{apert} can be straightforwardly worked out after an integration of the right-hand-sides of the Gauss equations {in \rfr{Gausseq}}, evaluated onto the unperturbed Keplerian ellipse  {where $\mu$ is to be assumed as constant}, over one orbital revolution  {by means of
\eqi dt=\left(\rp{1-e\cos E}{n}\right)dE; \eqf} we assume $\dot m/m$ constant over one orbital period. \textcolor{black}{If the mass variation occurred at fast rates with respect to the orbital frequency, it should explicitly be treated as a specific function of time in \rfr{equazza2} and \rfr{apert}. } They are
\begin{equation}
\left\{
\begin{array}{lll}
\dert a t & = & -2a\dmdt, \\ \\
\dert e t & = & 0, \\ \\
\dert I t & = & 0, \\ \\
\dert\Omega t & = & 0, \\ \\
\dert\omega t & = & \dmdt\rp{\vesc\sqrt{1-e^2}}{n a}, \\ \\
\dert{\mathcal{M}}t & = & -\dmdt\rp{3\vesc}{n a}.
\end{array}\lb{gauss}
\right.
\end{equation}
Notice that \rfr{gauss} are mathematically exact in the sense that no simplifying assumptions concerning $e$ were adopted in the calculation. On the other hand, from a physical point of view it is necessary to require moderate values for the eccentricity since, otherwise, it would not be possible to consider $\dot m/m$ constant over the integration performed over one full orbital revolution.  {It is worthwhile noticing that, even in the case of almost circular orbits, the long-term variations of \rfr{gauss} can be considered as secular changes only over  timescales in which it is possible to assume $\dot m/m$ as constant; otherwise,  slow time-dependent modulations, depending on the exact law of variation of $m=m(t)$, occur. Here we just recall that the long-term variations caused by an isotropic mass loss of the parent star were worked out with the same approach in \citet{Iorio}: it turned out that all the osculating Keplerian orbital elements remain unchanged, apart from the osculating semi-major axis and eccentricity which undergo  long-term variations
\eqi
\left\{
\begin{array}{lll}
\dot a_{\rm iso} &=& 2\left(\rp{e}{1-e}\right)\left(\rp{\dot M}{M}\right)a, \\ \\
\dot e_{\rm iso} &=& \left(\rp{\dot M}{M}\right)\left(1+e\right).
\end{array}\lb{bunga}
\right.
\eqf This outwardly counter-intuitive result, and the lack of actual contradiction with the occurring expansion\footnote{ {Indeed, from  eq. (27) and eq. (37) of \citet{Iorio} it can be inferred that the secular change of the star-planet distance is $d\Delta/dt\propto -\left(\dot M/M\right)a\left(1-e\right)$, so that the orbit actually expands for $\dot M/M < 0$, as expected.}} of the true orbit for a mass decrease, are fully discussed in \citet{Iorio} with abundance of explicative pictures. \textcolor{black}{The expressions of \rfr{bunga} were also used  by \citet{Ioriob} in the framework of dark matter studies in our solar system}. In any case, we notice that, given the order of magnitude of isotropic mass losses in typical Sun-like main sequence stars (cfr. with \rfr{pazzia}), such effects are smaller than those investigated here by some orders of magnitude.  }

From \rfr{gauss} it turns out that the semi-major axis $a$ undergoes a temporal variation proportional to $\dot m/m$; it is independent of the eccentricity and of $V_{\rm esc}$. The eccentricity is, instead, left unaffected. As expected, if $\dot m<0$ the size of the orbit increases since $\dot a>0$.

We qualitatively checked our analytical results by numerically integrating the equations of motion  {of \rfr{equazza2}}, in cartesian coordinates, for a fictitious hot Jupiter experiencing a constant mass loss during its motion along a Sun-like parent star, and tracing the osculating Keplerian ellipses at two consecutive pericenter passages. The results are displayed in Figure \ref{figura1} for an almost circular initial orbit, and in Figure \ref{figura2} for a highly elliptical initial orbit; they have the unique purpose of effectively displaying the mathematical agreement with the analytical results, and the percent mass loss was purposely greatly exaggerated just to this aim.  {Given the merely illustrative purpose of the figures displayed here,  just aimed to numerically confirm the qualitative features of the effects of \rfr{gauss}, longer time intervals for the numerical integrations are unnecessary.}
\begin{figure}
\begin{center}
\includegraphics[width=10cm,angle=0]{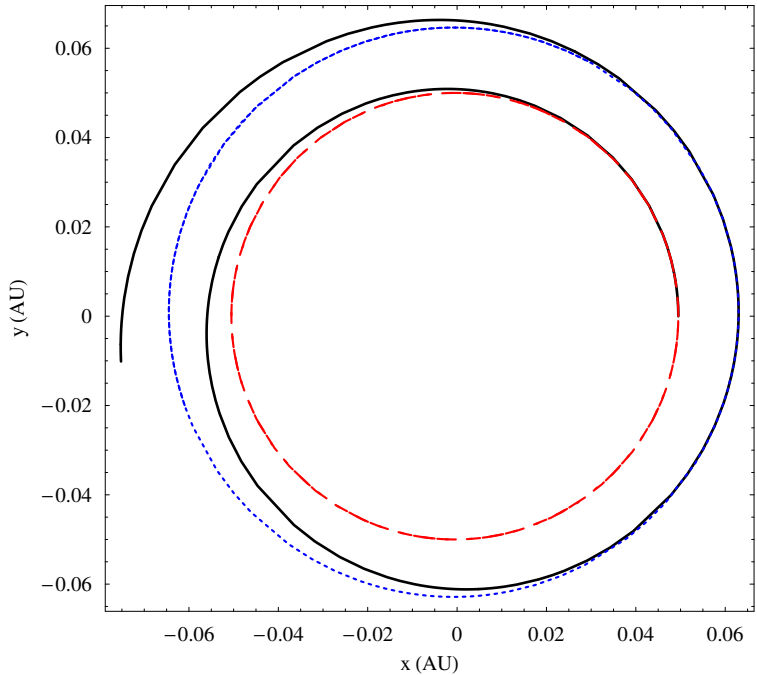}
\caption{Black continuous line: numerically integrated trajectory of a fictitious Jovian-type planet experiencing a purposely exaggerated constant percent mass loss $\dot m/m=-0.025$ d$^{-1}$, with $V_{\rm esc}=10^4$ m s$^{-1}$, in the field of a Sun-like star. The initial conditions are  $x_0=a_0(1-e_0),y_0=0,z_0=0,\dot x_0=0,\dot y_0 = n_0 a_0\sqrt{\rp{1+e_0}{1-e_0}},\dot z_0=0$, with $a_0=0.05$ au, $e_0=0.01$, $P_{\rm b}\doteq 2\pi/n_0=4.08$ d. The numerical integration is over $\Delta t=2 P_{\rm b}=8.16$ d. Red dashed line: osculating Keplerian ellipse at the first pericenter passage $(t=0)$. Blue dotted line: osculating Keplerian ellipse at the second pericenter passage ($t=4.85$ d). It can be clearly noticed that the eccentricities of both the osculating Keplerian ellipses remain constant, contrary to their semi-major axes which, instead, increase.}\lb{figura1}
\end{center}
\end{figure}
\begin{figure}
\begin{center}
\includegraphics[width=10cm,angle=0]{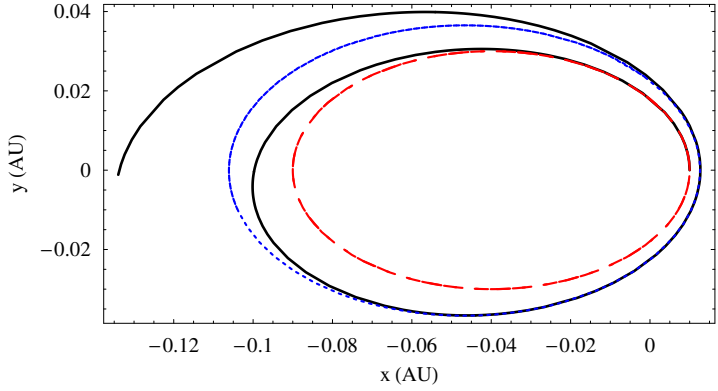}
\caption{Black continuous line: numerically integrated trajectory of a fictitious Jovian-type planet experiencing a purposely exaggerated constant percent mass loss $\dot m/m=-0.025$ d$^{-1}$, with $V_{\rm esc}=10^4$ m s$^{-1}$, in the field of a Sun-like star. The initial conditions are  $x_0=a_0(1-e_0),y_0=0,z_0=0,\dot x_0=0,\dot y_0 = n_0 a_0\sqrt{\rp{1+e_0}{1-e_0}},\dot z_0=0$, with $a_0=0.05$ au, $e_0=0.8$, $P_{\rm b}\doteq 2\pi/n_0=4.08$ d. The numerical integration is over $\Delta t=2 P_{\rm b}=8.16$ d. Red dashed line: osculating Keplerian ellipse at the first pericenter passage $(t=0)$. Blue dotted line: osculating Keplerian ellipse at the second pericenter passage ($t=4.53$ d). It can be clearly noticed that the eccentricities of both the osculating Keplerian ellipses remain constant, contrary to their semi-major axes which, instead, increase.}\lb{figura2}
\end{center}
\end{figure}
The increase of the semi-major axis and the constancy of the eccentricity of the osculating Keplerian ellipses are apparent.  {Analogous pictures for the isotropic mass loss case  can be found in \citet{Iorio}.}
Also from a quantitative point of view the agreement with out analytical results is excellent. Indeed, in Figure \ref{figura3} we depict the change of the semi-major axis over a Keplerian orbital period obtained from a numerical integration of the equations of motion of a fictitious hot Jupiter  in cartesian coordinates with and without a constant mass-loss of\footnote{In order to make a meaningful comparison with our analytical, perturbative results we choose a value for $|\dot m/m|$ small enough to make the extra-acceleration of \rfr{apert} much smaller (5 orders of magnitude) than the Newtonian monopole over the interval of the numerical integration.}   $\dot m/m=-0.00025$ d$^{-1}$; both the integrations use $a_0=0.05$ au and $e_0=0.1$. The overall shift is just equal to the one which can be inferred from \rfr{gauss}.
\begin{figure}
\begin{center}
\includegraphics[width=10cm,angle=0]{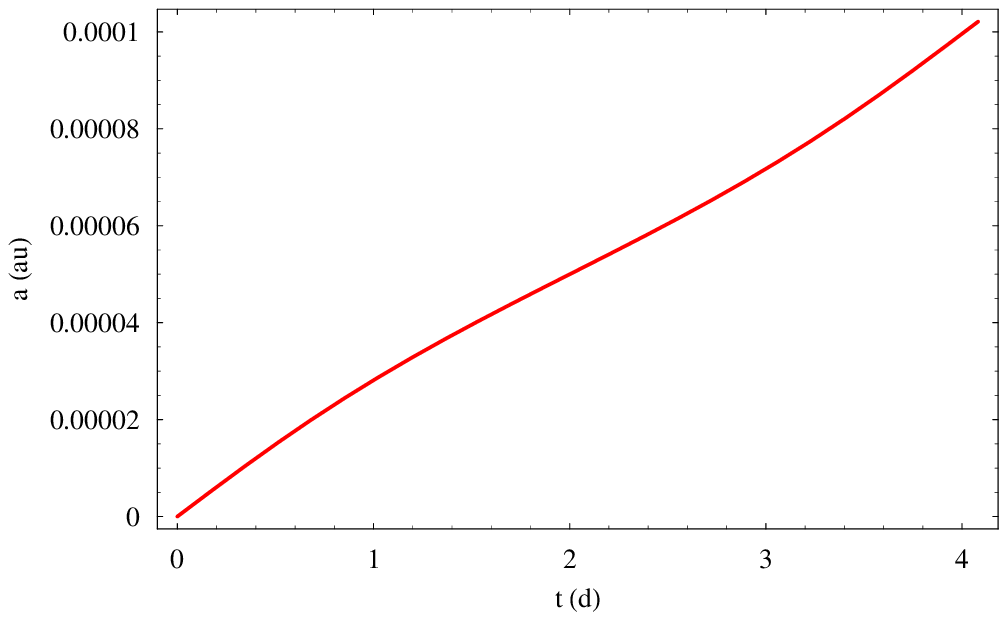}
\caption{Difference between the semi-major axes of a fictitious Jovian-type planet in the field of a Sun-like star computed from the numerically integrated equations of motion  {of \rfr{equazza2}} with and without a constant percent mass loss  $\dot m/m=-0.00025$ d$^{-1}$ and  $V_{\rm esc}=10^4$ m s$^{-1}$. The initial conditions are  $x_0=a_0(1-e_0),y_0=0,z_0=0,\dot x_0=0,\dot y_0 = n_0 a_0\sqrt{\rp{1+e_0}{1-e_0}},\dot z_0=0$, with $a_0=0.05$ au, $e_0=0.1$, $P_{\rm b}\doteq 2\pi/n_0=4.08$ d. The numerical integration is over $\Delta t=P_{\rm b}=4.08$ d. The net increase of $a$ is apparent, with an overall magnitude equal to the one which can be inferred from $\dot a=-2a (\dot m/m)$ of \rfr{gauss} obtained perturbatively.}\lb{figura3}
\end{center}
\end{figure}
Figure \ref{figura4}, obtained for $e_0=0.8$, confirms the analytical finding of \rfr{gauss} that the net change in $a$ is, actually, independent of $e$.
\begin{figure}
\begin{center}
\includegraphics[width=10cm,angle=0]{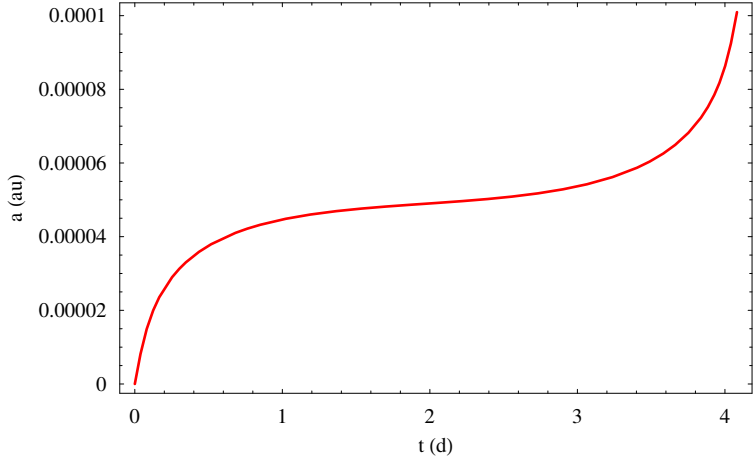}
\caption{Difference between the semi-major axes of a fictitious Jovian-type planet in the field of a Sun-like star computed from the numerically integrated equations of motion  {of \rfr{equazza2}} with and without a constant percent mass loss  $\dot m/m=-0.00025$ d$^{-1}$ and  $V_{\rm esc}=10^4$ m s$^{-1}$. The initial conditions are  $x_0=a_0(1-e_0),y_0=0,z_0=0,\dot x_0=0,\dot y_0 = n_0 a_0\sqrt{\rp{1+e_0}{1-e_0}},\dot z_0=0$, with $a_0=0.05$ au, $e_0=0.8$, $P_{\rm b}\doteq 2\pi/n_0=4.08$ d. The numerical integration is over $\Delta t=P_{\rm b}=4.08$ d. The net increase of $a$ is apparent, with an overall magnitude equal to the one which can be inferred from $\dot a=-2a (\dot m/m)$ of \rfr{gauss} obtained perturbatively. Cfr. with Figure \ref{figura3}.}\lb{figura4}
\end{center}
\end{figure}
In Figure \ref{figura5} we show that also the numerically integrated variation of the eccentricity does not exhibit any secular trend, in agreement with \rfr{gauss}.
\begin{figure}
\begin{center}
\includegraphics[width=10cm,angle=0]{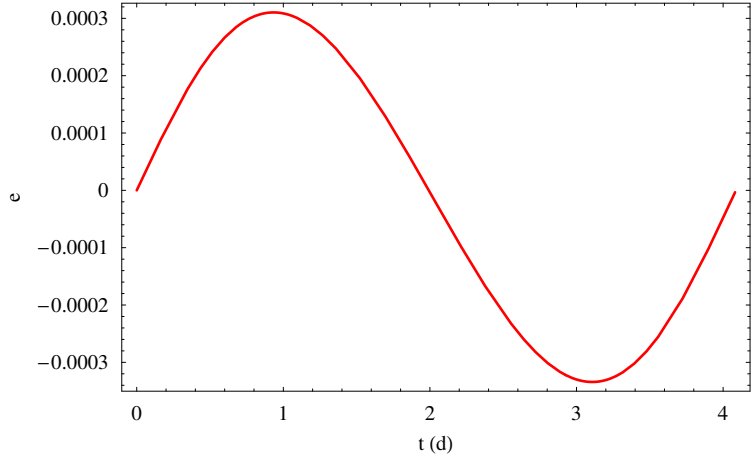}
\caption{Difference between the eccentricities of a fictitious Jovian-type planet in the field of a Sun-like star computed from the numerically integrated equations of motion  {of \rfr{equazza2}} with and without a constant percent mass loss  $\dot m/m=-0.00025$ d$^{-1}$ and  $V_{\rm esc}=10^4$ m s$^{-1}$. The initial conditions are  $x_0=a_0(1-e_0),y_0=0,z_0=0,\dot x_0=0,\dot y_0 = n_0 a_0\sqrt{\rp{1+e_0}{1-e_0}},\dot z_0=0$, with $a_0=0.05$ au, $e_0=0.1$, $P_{\rm b}\doteq 2\pi/n_0=4.08$ d. The numerical integration is over $\Delta t=P_{\rm b}=4.08$ d. The absence of a net increase of $e$ is apparent, in agreement with $\dot e=0$ of \rfr{gauss} obtained perturbatively.}\lb{figura5}
\end{center}
\end{figure}
According to \rfr{gauss}, the pericenter and the mean anomaly experience secular precessions which depend on $V_{\rm esc}$ and on $e$. Such a behavior  is qualitatively confirmed by Figure \ref{figura6} displaying the result of a numerical integration of a fictitious hot Jupiter obtained  for suitably chosen values of $\dot m/m,V_{\rm esc},e_0$; once again, it has just illustrative purposes.
\begin{figure}
\begin{center}
\includegraphics[width=10cm,angle=0]{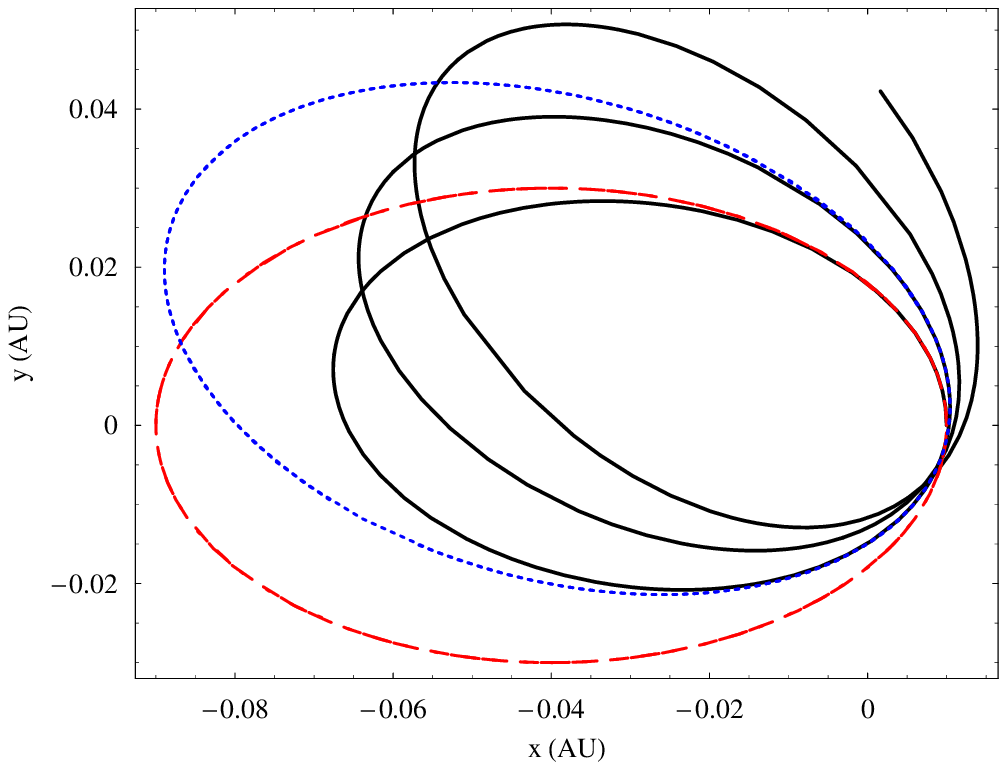}
\caption{Black continuous line: numerically integrated trajectory of a fictitious Jovian-type planet experiencing a purposely exaggerated constant percent mass loss $\dot m/m=-0.0025$ d$^{-1}$, with $V_{\rm esc}=10^7$ m s$^{-1}$, in the field of a Sun-like star. The initial conditions are  $x_0=a_0(1-e_0),y_0=0,z_0=0,\dot x_0=0,\dot y_0 = n_0 a_0\sqrt{\rp{1+e_0}{1-e_0}},\dot z_0=0$, with $a_0=0.05$ au, $e_0=0.8$, $P_{\rm b}\doteq 2\pi/n_0=4.08$ d. The numerical integration is over $\Delta t=2 P_{\rm b}=8.16$ d. Red dashed line: osculating Keplerian ellipse at the first pericenter passage $(t=0)$. Blue dotted line: osculating Keplerian ellipse at the second pericenter passage ($t=2.56$ d). The retrograde precession of the pericenter of the osculating Keplerian ellipses is apparent.}\lb{figura6}
\end{center}
\end{figure}

Our result about the semi-major axis is, in principle, important since it yields a physical mechanism which counteracts the inward migration of the exoplanet after it reaches a distances small enough to trigger an effective mass depletion. The typical figures previously obtained yields
\eqi \dot a \sim 2.5\ {\rm m\ yr^{-1}},\eqf i.e. an orbit with $a=0.05$ au increases its size by $2.5$ m after each year if the planet following it loses mass at a percent rate of about $|\dot m/m|\sim 10^{-10}$ yr$^{-1}$; by postulating that $\dot m/m$ remains almost constant over the aeons, we would have
\eqi\Delta a \sim 1.1\times 10^{10}\ {\rm m}=0.075\ {\rm au}\eqf after a time span $\Delta t = 4.5$ Gyr. Thus, the overall orbital evolution of close exoplanets may be affected by their mass depletion. Concerning the retrograde pericenter precession of \rfr{gauss}, it may, in principle, be viewed as a potential source of systematic bias in some proposed attempts to measure the general relativistic pericenter precession \citep{peri1,peri2,peri3,peri4}; anyway,  it turns out to be smaller by several orders of magnitude than the relativistic one for the typical values of the relevant parameters adopted so far.
\section{Summary and conclusions}\lb{tre}
We  investigated the orbital consequences of a non-isotropic mass depletion affecting the evaporating atmosphere of a typical hot Jupiter in a close ($a=0.05$ au) orbit along a Sun-like parent star.

Analytical perturbative calculation showed that the semi-major axis undergoes a  {long-term} increment which is independent of both the escape velocity of the atmosphere and of the eccentricity of the orbit, which, instead, remains constant. Such results were confirmed, both qualitatively and quantitatively, by numerical integrations of the equations of motion in cartesian coordinates. By assuming $|\dot m|\lesssim 10^{17}$ kg yr$^{-1}$,  {corresponding to $|\dot m/m|\lesssim 10^{-10}$ yr$^{-1}$ for a Jovian mass}, it turns out that an increase of a few meters per year occurs for the semi-major axis. Such a mechanism may be important for the dynamical evolution of hot Jupiters, especially in connection with the issue of their inward migrations which should bring them close to their hosting stars from the supposedly remote regions of their formation. Indeed, it may explain, at least in principle, why such a phenomenon seems to cease after the planet reaches a distance of not less than $0.01$ au.  {It is just the case to notice that such a behavior would be enhanced, in principle, by the isotropic mass variation $\dot M$ of the hosting star as well, although it may be some orders of magnitude smaller for typical Sun-like values of $\left.\dot M/M\right|_{\rm iso}$.
}

The pericenter of the orbit of the evaporating planet experiences a  {long-term} retrograde precession which, however, is not a concern for some proposed detections of the general relativistic pericenter precession since it is several orders of magnitude smaller.
\section*{Acknowledgements}
I thank D. Ehrenreich for interesting correspondence and L.-S. Li for having provided me with useful references.
I am also grateful to an anonymous referee for her/his useful comments.



\begin{thebibliography}{}


\bibitem[\protect\citeauthoryear{Armellini}{1935}]{armel}
Armellini, G., 1935. The Observatory, 58, 158.


\bibitem[\protect\citeauthoryear{Bouchy et al.}{2005}]{Bou05}	
Bouchy, F., Udry, S., Mayor, M., Moutou, C., Pont, F., Iribarne, N., da Silva, R., Ilovaisky, S., Queloz, D., Santos, N.C., S\'{e}gransan, D., Zucker, S., 2005. Astron. Astrophys., 444, L15.

\bibitem[\protect\citeauthoryear{Brouwer and  Clemence}{1961}]{Brou}
Brouwer, D., Clemence, G.M., 1961. Methods of celestial mechanics, Academic Press, New York.

\bibitem[\protect\citeauthoryear{Charbonneau et al.}{2000}]{Char00}	
Charbonneau, D., Brown, T.M., Latham, D.W., Mayor, M., 2000. Astroph. J., 529, L45.


\bibitem[\protect\citeauthoryear{Deprit}{1983}]{deprit}
Deprit, A., 1983. Celest. Mech. Dyn. Astron., 31, 1.


\bibitem[\protect\citeauthoryear{Ehrenreich et al.}{2008}]{eren}
Ehrenreich, D., Lecavelier Des Etangs, A., H\'{e}brard, G., D\'{e}sert, J.-M., Vidal-Madjar, A., McConnell, J.C., Parkinson, C.D., Ballester, G.E., Ferlet, R., 2008.
Astron. Astrophys., 483, 933.

\bibitem[\protect\citeauthoryear{Ehrenreich and  D\'{e}sert}{2011}]{trans}
Ehrenreich, D., D\'{e}sert, J.-M., 2011. Astron. Astrophys., 529, A136.

\bibitem[\protect\citeauthoryear{Fossati et al.}{2010}]{Foss}
Fossati, L., Haswell, C.A., Froning, C.S., Hebb, L.,  Holmes, S.,  Kolb, U.,  Helling, Ch.,  Carter, A.,  Wheatley, P.,  Cameron, A.C.,  Loeillet, B.,  Pollacco, D.,  Street, R.,  Stempels, H.C.,  Simpson, E.,  Udry, S.,  Joshi, Y.C.,  West, R.G.,  Skillen, I.,  Wilson, D., 2010. Astroph. J., 714, L222.


\bibitem[\protect\citeauthoryear{Gylden}{1884}]{gyl}
Gylden, H., 1884. Astron. Nachr., 109, 1.


\bibitem[\protect\citeauthoryear{Hadjidemetriou}{1963}]{Hadji1}
Hadjidemetriou, J.D., 1963. Icarus, 2, 440.


\bibitem[\protect\citeauthoryear{Hadjidemetriou}{1966}]{had}
Hadjidemetriou, J.D., 1966. Icarus,  5, 34.


\bibitem[\protect\citeauthoryear{Hadjidemetriou}{1967}]{Hadji}
Hadjidemetriou, J.D., 1967. Adv. Astron. Astrophys., 5, 131.

\bibitem[\protect\citeauthoryear{Hebb et al.}{2009}]{Hebb09}
Hebb, L., Collier-Cameron, A., Loeillet, B., Pollacco, D., H\'{e}brard, G., Street, R.A., Bouchy, F., Stempels, H.C., Moutou, C., Simpson, E., Udry, S., Joshi, Y.C., West, R.G., Skillen, I., Wilson, D.M., McDonald, I., Gibson, N.P., Aigrain, S., Anderson, D.R., Benn, C.R., Christian, D.J., Enoch, B., Haswell, C.A., Hellier, C., Horne, K., Irwin, J., Lister, T.A., Maxted, P., Mayor, M., Norton, A.J., Parley, N., Pont, F., Queloz, D., Smalley, B., Wheatley, P.J., 2009. Astroph. J., 693, 1920.

\bibitem[\protect\citeauthoryear{Heyl and  Gladman}{2007}]{peri2}
Heyl, J.S., Gladman, B.J., 2007. Mon. Not. Roy. Astron. Soc., 377, 1511.

\bibitem[\protect\citeauthoryear{Henry et al.}{2000}]{Hen00}	
Henry, G.W., Marcy, G.W., Butler, R.P., Vogt, S.S., 2000. Astroph. J., 529, L41.


\bibitem[\protect\citeauthoryear{Iorio}{2010a}]{Iorio}
Iorio, L., 2010a. Scholarly Research Exchange, 2010, 261249.

\bibitem[\protect\citeauthoryear{Iorio}{2010b}]{Ioriob}
\textcolor{black}{Iorio, L., 2010b. J. Cosmol. Astropart. Phys., 05, 018.}

\bibitem[\protect\citeauthoryear{Iro and  Deming}{2010}]{iro}
Iro, N.,  Deming, L.D., 2010. Astroph. J., 712, 218.


\bibitem[\protect\citeauthoryear{Jeans}{1924}]{jeans}
Jeans, J.H., 1924. Mon. Not. Roy. Astron. Soc., 85,  2.



\bibitem[\protect\citeauthoryear{Jeans}{1961}]{jeans61}
Jeans, J.H., 1961. Astronomy and Cosmogony, Dover, New York.


\bibitem[\protect\citeauthoryear{Jord\'{a}n and  Bakos}{2008}]{peri3}
Jord\'{a}n, A., Bakos, G., 2008. Astroph. J., 685, 543.


\bibitem[\protect\citeauthoryear{Kevorkian and  Cole}{1966}]{kevo}
Kevorkian, J.,  Cole, J.D., 1966. Multiple Scale and Singular
Perturbation Methods, Springer, New York.



\bibitem[\protect\citeauthoryear{Kholshevnikov and   Fracassini}{1968}]{khol}	
Kholshevnikov, K.V.,  Fracassini, M., 1968. Conferenze
dell' Osservatorio Astronomico di Milano-Merate, Serie I, no. 9,
pp. 5–50.




\bibitem[\protect\citeauthoryear{Kuryshev and   Perov}{1981}]{kury}		
Kuryshev, V.I., Perov, N.I., 1981. Soviet Astronomy, 25, 504.



\bibitem[\protect\citeauthoryear{Lecavelier Des Etangs et al.}{2010}]{Leca}		
Lecavelier Des Etangs, A., Ehrenreich, D., Vidal-Madjar, A., Ballester, G.E., D\'{e}sert, J.-M., Ferlet, R., H\'{e}brard, G., Sing, D.K., Tchakoumegni, K.-O., Udry, S., 2010. Astron. Astrophys., 514, A72.


\bibitem[\protect\citeauthoryear{Li et al.}{2003}]{Li03}	
Li, L.-S., Yu, L.-Z., Zheng, X.-T., 2003. Publications of the Yunnan Observatory, 97,  1.



\bibitem[\protect\citeauthoryear{Li}{2008}]{Li08}	
Li, L.-S., 2008. Astronomy Reports, 52,  806.

\bibitem[\protect\citeauthoryear{Li}{2009}]{Li09}
Li, L.-S., 2009. Int. J. Mod. Phys. D, 18,  1243.

\bibitem[\protect\citeauthoryear{Lin et al.}{1996}]{Lin96}
Lin, D.N.C., Bodenheimer, P., Richardson, D.C., 1996. Nature, 380, 606.




\bibitem[\protect\citeauthoryear{Linsky et al.}{2010}]{Lins}	
Linsky, J.L., Yang, H., France, K., Froning, C.S., Green, J.C., Stocke, J.T., Osterman, S.N., 2010. Astroph. J., 717, 1291.

\bibitem[\protect\citeauthoryear{Lubow and  Ida}{2010}]{Lubo}	
Lubow, S.H., Ida, S., 2010. Planet Migration, In: Seager S. (ed.) \textit{Exoplanets}, Arizona University Press, Tucson, pp. 347-371.

\bibitem[\protect\citeauthoryear{Me\v{s}\v{c}erskii}{1897}]{Mesc}
Me\v{s}\v{c}erskii, I.V., 1897. Mass Variable Particle Dynamics, St. Petersburg.

\bibitem[\protect\citeauthoryear{Miralda-Escud\'{e}}{2002}]{peri1}
Miralda-Escud\'{e}, J., 2002. Astroph. J., 564, 1019.

\bibitem[\protect\citeauthoryear{Naef et al.}{2001}]{naef}	
Naef, D.,  Latham, D. W.,  Mayor, M.,  Mazeh, T.,  Beuzit, J. L.,  Drukier, G. A.,  Perrier-Bellet, C.,  Queloz, D.,  Sivan, J. P.,  Torres, G.,  Udry, S.,  Zucker, S., 2001. Astron. Astrophys., 375, 27.	

\bibitem[\protect\citeauthoryear{Omarov}{1962}]{omarov62}
\textcolor{black}{Omarov, T.B., 1962. Izv. Astrofiz. Inst. AN Kazakh SSR, 14, 66.}

\bibitem[\protect\citeauthoryear{Omarov}{1964}]{omarov}
Omarov, T.B., 1964. Soviet Astronomy, 8, 127.


\bibitem[\protect\citeauthoryear{P\'{a}l and  Kocsis}{2008}]{peri4}	
P\'{a}l, A., Kocsis, B., 2008. Mon. Not. Roy. Astron. Soc., 389, 191.

\bibitem[\protect\citeauthoryear{Plastino and  Muzzio}{1992}]{erro}
Plastino, A.R., Muzzio, J.C., 1992. Celest. Mech. Dyn. Astron., 53, 227.


\bibitem[\protect\citeauthoryear{Polyakhova}{1989}]{polya}
Polyakhova, E.N., 1989. Soviet Astronomy, 33, 673.



\bibitem[\protect\citeauthoryear{Prieto and  Docobo}{1997a}]{prietoa}
Prieto, C., Docobo, J.A., 1997a. Astron. Astrophys., 318, 657.

\bibitem[\protect\citeauthoryear{Prieto and  Docobo}{1997b}]{prietob}
\textcolor{black}{Prieto, C., Docobo, J.A., 1997b. Celest. Mech. Dyn. Astron., 68, 53.}

\bibitem[\protect\citeauthoryear{Rahoma et al.}{2009}]{rahoma}
\textcolor{black}{Rahoma, W.A., Abd El-Salam, F.A., Ahmed, M.K., 2009, J. Astrophys. Astr., 30, 187.}

\bibitem[\protect\citeauthoryear{Razbitnaya}{1985}]{russa}	
Razbitnaya, E.P., 1985. Soviet Astronomy 29, 684.


\bibitem[\protect\citeauthoryear{Schr\"{o}der and  Smith}{2008}]{MASLOS}
Schr\"{o}der, K.P., Smith, R.C., 2008.  Mon. Not.  Roy. Astron. Soc., 386, 155.

\bibitem[\protect\citeauthoryear{Sommerfeld}{1952}]{Somm}
Sommerfeld, A., 1952. Mechanics. Lectures on Theoretical Physics, Vol. I., New York, p. 28.


\bibitem[\protect\citeauthoryear{Str\"{o}mgren}{1903}]{strom}
Str\"{o}mgren, E., 1903.  Astron.
Nachr., 163, 129.



\bibitem[\protect\citeauthoryear{Verhulst and  Eckhaus}{1970}]{verh}
Verhulst, F.,  Eckhaus, W., 1970. Int. J. of Non-Linear Mechanics, 5, 617.



\bibitem[\protect\citeauthoryear{Vescan}{1937}]{vescan}
Vescan, T.T., 1937.
Bull. scient. de l'\'{E}cole Polyt. Timi\c{s}oara, 7, fasc. 1-2.



\bibitem[\protect\citeauthoryear{Vescan}{1939}]{vescan2}
Vescan, T.T., 1939.
Contribu\c{t}ii la teoria cinetic\u{a} \c{s}i relativist\u{a} a fluidelor reale, Ph. D. Thesis, Cluj.


\bibitem[\protect\citeauthoryear{Vidal-Madjar et al.}{2003}]{vida03}
Vidal-Madjar, A.,  Lecavelier des Etangs, A.,  D\'{e}sert, J.-M.,  Ballester, G.E.,  Ferlet, R.,  H\'{e}brard, G.,  Mayor, M., 2003. Nature, 422,  143.

\bibitem[\protect\citeauthoryear{Vidal-Madjar et al.}{2004}]{vida04}
Vidal-Madjar, A., D\'{e}sert, J.-M., Lecavelier des Etangs, A., H\'{e}brard, G., Ballester, G.E., Ehrenreich, D., Ferlet, R., McConnell, J.C., Mayor, M., Parkinson, C.D., 2004. Astroph. J., 604, L69.

\bibitem[\protect\citeauthoryear{Vidal-Madjar et al.}{2008}]{vida08}	
Vidal-Madjar, A., Lecavelier des Etangs, A., D\'{e}sert, J.-M., Ballester, G.E., Ferlet, R., H\'{e}brard, G., Mayor, M., 2008. Astroph. J.,  676,  L57.



\end{thebibliography}
\end{document}